# Vectorial modeling of near-field imaging with uncoated fiber probes: Transfer function and resolving power

**Niels Gregersen and Bjarne Tromborg**

Department of Communications, Optics & Materials, Nano•DTU,

Technical University of Denmark, Building 345V, DK-2800 Kgs. Lyngby, Denmark

**Sergey I. Bozhevolnyi**

Department of Physics and Nanotechnology, Aalborg University,

Skjernvej 4A, DK-9220 Aalborg Øst, Denmark

**Abstract**

Using exact three-dimensional vectorial simulations of radiation coupling into uncoated dielectric fiber probes, we calculate amplitude transfer functions for conical single-mode fiber tips at the light wavelength of 633 nm. The coupling efficiency of glass fiber tips is determined in a wide range of spatial frequencies of the incident radiation for opening angles varying from 30° to 120°. The resolution in near-field imaging with these tips is considered for field distributions limited in both direct and spatial-frequency space. The characteristics of the transfer functions describing the relation between probed optical fields and near-field images are analyzed in detail. The importance of utilizing a perfectly sharp tip is also examined.

# 1. Introduction

The invention[1,2] of the scanning near-field optical microscope (SNOM) has opened new avenues in optical microscopy. Previously, the detection of scattered light from a sample under study took place in the far-field only, limiting the achievable resolution to approximately half the wavelength[3]. The SNOM overcomes this problem by measuring the field at a distance of only a few tens of nanometers above the sample[4], close enough to detect sub-wavelength components of the optical field; the near-field, before these decay below the background radiation floor. The SNOM has successfully been used in a wide range of near-field detection experiments; i.e. to achieve image resolution of ~ $\lambda/40$ at visible wavelengths[5], to study the excitation of surface plasmon polaritons[6] and to image light propagation in photonic crystal waveguides[7].

A commonly used configuration is the collection-mode SNOM[4]. It features a nano-collector consisting of a single-mode fiber terminating in a tapered tip. The tip is scanning above the sample and couples the near-field to the guided mode of the fiber. The propagating signal is detected remotely and a SNOM image is obtained.

The interpretation of the measured image, however, has been subject to debate[8-13]. The SNOM clearly allows detection of sub-wavelength features, but in spite of progress in the understanding of the field-image coupling mechanism, the exact relation between the optical field and the observed SNOM image remains unclear.

Near-field detection is generically complicated by the close proximity of the fiber tip, resulting in tip-sample interaction modifying the free-space optical field generated in absence of the nano-collector. The detected field thus includes a perturbation of the original free-space field, a perturbation that is generally difficult or impossible to separate. Fortunately, it has been shown that for a dielectric sample studied in the photon tunneling SNOM setup, the tip-sample interaction is negligible[14], and in this case the fiber tip can be approximated to a passive detector of the optical field.

Initially, the nano-collector was considered a point-like probe[8] detecting the optical field intensity at a specific position. It was argued[9] that the finite size of the probe should be taken into account by introducing a transfer function (TF). The intensity transfer function (ITF), relating the absolute square of the electric field to the SNOM image, was proposed[9] and the conception that the image represents field intensity below a spatial frequency cut-off has been used to analyze experimental near-field data[15,16].

It was then shown that the ITF does not accurately model the field-image relation[10]. The correct general framework necessary for describing the field-image coupling was given in Ref. 11. It was pointed out[12] that when utilizing a single-mode fiber, the general TF can be described by an amplitude coupling function. General symmetry properties of this TF, correctly relating the optical field to the image, are given in Ref. 13.

Further information about the nature of the TF was obtained as comparisons[17-19] between modeling and experiment revealed correspondence between detected SNOM images and the intensities of the free space fields in planes a distance above the tip apex. This hints that the optical field may be coupled to the guided fiber mode not at the very tip apex, but in an effective detection point inside the tip.

However, to completely understand the field-image relation, exact numerical modeling of the field scattered on the fiber tip must be performed. A comparison between free-space field and image measured using a reduced 2D tip without invoking the concept of TF was performed in Ref. 20, and calculations of 2D TFs were presented in Ref. 21. We present here a technique to numerically determine TFs for realistic uncoated fiber tips using full 3D vectorial modeling and we have determined the TFs for conical tips of varying geometry. The main results are that optimal field to fiber coupling is obtained for a large rather than a small opening angle, and that the fiber tip need not be sharp.

This article is organized as follows: In section 2 the theoretical framework necessary to describe the coupling of the optical field is summarized. We describe the modeling technique employed in section 3. The results of the numerical simulations are given in section 4, and the implications for the field-image relation are discussed in section 5. A conclusion follows in section 6.

## 2. Near-field coupling

The general experimental SNOM setup we will model is illustrated in Fig. 1. A dielectric sample with sub-wavelength features is placed on a prism and is subject to illumination from below at the wavelength $\lambda$ of 633 nm. The sample scatters the light, and a near-field is generated.

The SNOM fiber tip is scanned above the sample in constant height mode[4] with tip apex in a detection plane $z = z_d$. Other scanning modes, such as constant distance mode[4], are frequently used in SNOM microscopy, but the TF relates the field in a detection plane of fixed $z$ coordinate to the image and the choice of constant height mode is thus necessary if field-image comparisons are to be made.

The tip couples the field to the guided fiber mode but also scatters light back towards the sample. We restrict our analysis to SNOM measurements on weakly reflecting dielectric samples where the tip-sample interaction can be ignored[14] and we thus only treat the coupling of the near-field into the guided mode. Also, we assume that power levels are sufficiently low so that linear scattering can be assumed. These restrictions are made as the TF, relating the free-space field with the observed SNOM image, is only defined[11] in the linear scattering regime and in absence of multiple reflections between tip and sample. A treatment of the field-image relation in setups where the tip-sample interaction or non-linear effects are important is beyond the scope of this paper.

In the detection plane $(\mathbf{r}_\perp, z_d)$ we consider the electric field $\mathbf{E} = (E_x, E_y, E_z)$. We then make the plane-wave expansion in the detection plane:

$$\mathbf{E}(\mathbf{r}_\perp, z_d) = \int \mathbf{F}(\mathbf{k}_\perp) \exp(i\mathbf{k}_\perp \cdot \mathbf{r}_\perp) d\mathbf{k}_\perp. \tag{1}$$

The $z$-dependence of the plane waves are given by $\exp(ik_z z)$ with $k_z = \sqrt{k_0^2 - |\mathbf{k}_\perp|^2}$, where $k_0 = 2\pi/\lambda$. Plane waves with $|\mathbf{k}_\perp|^2 < k_0^2$ correspond to the propagating part of the near-field, while those with $|\mathbf{k}_\perp|^2 > k_0^2$ represent its evanescent part.

The three components of the vector field $\mathbf{E} = (E_x, E_y, E_z)$ are not independent but subject to the divergence relation $\nabla \cdot \mathbf{E} = 0$, valid in free space. For this reason we will in the following only consider the transverse components $\mathbf{E}_\perp = (E_x, E_y)$ and $\mathbf{F}_\perp = (F_x, F_y)$ of the field and of the expansion coefficients respectively. The $z$-component of the field can be determined at any time using

$$E_z(\mathbf{r}_\perp, z_d) = -\int \frac{\mathbf{F}_\perp \cdot \mathbf{k}_\perp}{k_z} \exp(i\mathbf{k}_\perp \cdot \mathbf{r}_\perp) d\mathbf{k}_\perp. \tag{2}$$

Now, the optical field couples to guided fiber modes $n$ with angular momenta $m$ with normalized mode profiles $\mathbf{G}^{n,m}(r) \exp(im\varphi)$. Following Ref. 11, we introduce vector coupling coefficients $\mathbf{H}^{n,m}(\mathbf{k}_\perp)$ describing the coupling between a plane wave with wave vector $\mathbf{k}_\perp$ and the guided fiber mode with field profile of the form $A^{n,m}\mathbf{G}^{n,m}(r)\exp(im\varphi)$. The contribution from the total field to a guided mode is then

$$A^{n,m}(\mathbf{r}_\perp) = \int \mathbf{H}^{n,m}(\mathbf{k}_\perp) \cdot \mathbf{F}_\perp(\mathbf{k}_\perp) \exp(i\mathbf{k}_\perp \cdot \mathbf{r}_\perp) d\mathbf{k}_\perp , \qquad (3)$$

where $\mathbf{r}_\perp$ refers to the position of the apex in the plane $z = z_d$. The mode profiles $\mathbf{G}^{n,m}(r)\exp(im\varphi)$ are normalized such that the total power $S$ detected at the end of the fiber is given by $S(\mathbf{r}_\perp) = \sum_{n,m} |A^{n,m}(\mathbf{r}_\perp)|^2$.

To take advantage of the rotational symmetry of the optical fiber, the plane wave vector components of Eq. (1) may be expressed in cylindrical coordinates as $\mathbf{F}_\perp(k,\theta) = F_p(k,\theta)\mathbf{e}_k + F_s(k,\theta)\mathbf{e}_\theta$, corresponding to $s$- and $p$-polarized parts of the field. We obtain:

$$\mathbf{E}_\perp(\mathbf{r}_\perp, z_d) = \int \left( F_p(k,\theta)\mathbf{e}_k + F_s(k,\theta)\mathbf{e}_\theta \right) \exp(i\mathbf{k}_\perp \cdot \mathbf{r}_\perp) d\mathbf{k}_\perp . \qquad (4)$$

The cylindrical unit vectors are illustrated in Fig. 2. We remark in passing that when calculating the $z$-component of the electric field using Eq. (2), only the $p$-polarized part of the field gives a contribution.

For a single-mode fiber of weak refractive index contrast we only need to include coefficients $(A_x, A_y)$ of the two orthogonal linearly polarized modes of the form $A_{x,y}\mathbf{G}_{x,y}(\mathbf{r}_\perp)$ where the profiles $\mathbf{G}_{x,y}(\mathbf{r}_\perp)$ as previously are normalized such that the total fiber power is $S(\mathbf{r}_\perp) = |A_x(\mathbf{r}_\perp)|^2 + |A_y(\mathbf{r}_\perp)|^2$. The field contribution to the two modes is then given by:

$$\begin{bmatrix} A_x(\mathbf{r}_\perp) \\ A_y(\mathbf{r}_\perp) \end{bmatrix} = \int \begin{bmatrix} -\sin\theta & \cos\theta \\ \cos\theta & \sin\theta \end{bmatrix} \begin{bmatrix} H_s(k)F_s(k,\theta) \\ H_p(k)F_p(k,\theta) \end{bmatrix} \exp(i\mathbf{k}_\perp \cdot \mathbf{r}_\perp) d\mathbf{k}_\perp . \tag{5}$$

The detailed derivation of this equation is given in the Appendix. In Eq. (5), $H_s(k)$ and $H_p(k)$ represent the coupling of an incoming plane wave of *s*- and *p*-polarization respectively with in-plane wave vector of norm *k* to a guided fiber mode. Due to symmetry, the functions $H_s(k)$ and $H_p(k)$ are independent of the angle of incidence $\theta$.

The functions $H_s(k)$ and $H_p(k)$ are the TFs that describe the detection capability of a particular fiber tip that we aim at determining numerically.

## 3. Modeling

In general, exact 3D vectorial simulation of light scattering by nano-structures requires extensive computer power, but for the fiber tip geometry we can exploit the rotational symmetry of the fiber. In cylindrical coordinates we can write the electric field as a sum of contributions of different angular momenta:

$$\mathbf{E}(r,\phi,z) = \sum_m \mathbf{e}_m(r,z) e^{im\phi} . \tag{6}$$

In a rotationally symmetric geometry such as the optical fiber, the various contributions are not coupled, and it is only necessary to perform calculations for the angular momentum of the guided fiber mode. This effectively reduces the number of dimensions from 3 to 2, making exact 3D vectorial calculations feasible with only modest computing power.

The method chosen in this work to model the scattering of plane waves into the guided fiber mode is the eigenmode expansion technique (EET). The structure is divided into layers along an axis of propagation, here chosen as the *z*-axis. The field is expanded on the eigenmodes in each layer and the fields on each side of a layer interface are connected using the transfer matrix formalism. The general method is well described in Ref. 22. However, the eigenmodes need to be determined. These can be obtained analytically in a cylindrically symmetric structure by solving a well established transcendental equation[23].

The fiber tip geometry modeled in this work is that of a tapered single-mode fiber with core and cladding refractive indices of 1.4622 and 1.4572 respectively and core diameter of 4 μm. Real optical fibers feature a cladding diameter of ~ 125 μm, but since the guided mode is no longer affected by the cladding-air index step for a cladding diameter above ~ 16 μm, the cladding diameter is limited to that value to save computation time. The wavelength λ is 633 nm in all calculations.

As the EET treats layered structures, we model the conically shaped tip using a large number of thin layers, resulting in a saw-tooth profile. The number of layers must be ~ 500 to sufficiently smoothen out the saw-tooth profile and achieve convergence of the TF. Here, the EET technique has an advantage compared to other techniques, as the total computation time increases only linearly with the number of layers.

A weakness of the EET, however, is the necessity of enclosing the fiber tip geometry in a metal cylinder introducing reflections from the cylinder wall. Our calculations were performed for cylinder diameters of 20, 40, 60 and 80 μm. In general we saw only negligible influence of the cylinder diameter on the TF, however for the opening angle of 90°, a small oscillatory dependence on cylinder diameter was observed in the first part of the propagating regime ($k/k_0 < 0.5$) of the TF. In this region the TF for β = 90° is thus slightly affected by numerical uncertainty.

With 1000 eigenmodes and 512 layers along the *z*-axis the total computation time for a TF is ~ 50 hours on a 2.6 GHz desktop PC with 1 GB of RAM.

## 4. Results

The TF allows us to determine the power carried by the guided fiber mode for a given incoming light intensity as function of $k = |\mathbf{k}_\perp|$. The following TFs are calculated for an electric field amplitude of 27.5 kV/m in the detection plane corresponding to an illumination intensity of 1 µW/µm$^2$ for a propagating wave.

TFs determined for *s*- and *p*-polarization for various opening angles are depicted in Fig. 3. In the propagating regime ($k/k_0 < 1$) we observe an oscillatory behavior for all the TFs of both polarizations. The peaks in this part of the spectrum represent coupling resonances, the positions of which depend on geometry. In the evanescent regime we generally observe a monotonic near-exponential decrease except for *p*-polarization in the interval $1 < k/k_0 < 1.5$ when β = 60° or 90°. We also note that for $k/k_0 > 1$ the coupling efficiency as function of opening angle seems to be at a maximum for an opening angle β around 90°. However we stress that determination of the exact optimal opening angle requires a study of TFs for more than four opening angles.

We have previously[21] performed 2D TF calculations on the same fiber tip geometry and we can thus directly compare the amplitude of the TFs for β = 30°. A reasonable qualitative correspondence is observed indicating that even though 3D computations are necessary to obtain the exact TF, fast 2D calculations may be useful to reveal its general shape.

### *4.1. Resolving power*

One of the motivations for calculating a TF is that it allows us to determine the resolving power of a SNOM fiber tip. The definition of resolution, however, depends on whether the optical fields distributions are limited in real space or in $k$-space.

To introduce the concept of resolution for distributions limited in real space, we consider an electric field distribution whose $x$- or $z$-component in the detection plane is of the shape of a delta-function. In the first case the in-plane field distribution is given by $\mathbf{E}_\perp(\mathbf{r}_\perp, z_d) = \delta(\mathbf{r}_\perp)\mathbf{e}_x$, resulting in $F_x(\mathbf{k}_\perp) = 1/(2\pi)^2$ for the plane wave expansion coefficients. $\nabla \cdot \mathbf{E} = 0$ must be satisfied however, so the $z$-component of the electric field distribution is non-zero and can be determined using Eq. (2). Secondly, when the $z$-component of the field is a delta-function we have $E_z(\mathbf{r}_\perp, z_d) = \delta(\mathbf{r}_\perp)$ and $F_z(\mathbf{k}_\perp) = 1/(2\pi)^2$, and this time the requirement $\nabla \cdot \mathbf{E} = 0$ leads to a non-zero in-plane field. As $s$-polarized waves do not contribute to the $z$-component of the field, we set $F_s = 0$ and obtain the $p$-polarized in-plane field contributions from $F_p(\mathbf{k}_\perp) = -k_z F_z(\mathbf{k}_\perp)/|\mathbf{k}_\perp|$, a rotationally symmetric function.

The calculated SNOM images for the delta-function distribution along the $x$-axis are shown in Fig. 4a. For $\beta = 30°$ the SNOM image profile features a near-Gaussian shape, but for the larger opening angles we generally observe an oscillating decay along the $x$-axis. This is not very surprising as the near-exponential decay of the TFs for $k/k_0 > 1$ serves as an effective cut-off and we can thus roughly approximate the product $\mathbf{H}^{n,m}(\mathbf{k}_\perp) \cdot \mathbf{F}_\perp(\mathbf{k}_\perp)$ of Eq. (3) to a step function equal to zero for $k/k_0 > 1$. Making the Fourier transform we obtain a sinc function having the oscillating decay observed in Fig. 4a for opening angles of 60, 90 and 120 degrees.

We characterize the spatial extent of the imaged spots using the average deviation (AD) along the two axes given by $\int S(x,0)|x|dx / \int S(x,0)dx$ along the $x$-axis, and the determined ADs are presented in Table 1. The smallest AD is obtained for β = 60° along the *y*-axis, however this opening angle also features the largest AD along the *x*-axis due to the slow oscillating decay. The minimum AD along the *x*-axis is given for the β = 90° opening angle with an AD along the *y*-axis only slightly larger than that for β = 60°. This behavior reflects the flatness of the TFs: The TF for β = 90° is rather broad and falls off slowly resulting in a wide step function giving a narrow sinc-like spot. On the other hand, the TF for β = 120° decreases much faster, the step function is of shorter width and the resulting spot is wider.

Fig. 4b presents the SNOM images obtained when the *z*-component of the field is a delta-function. The images preserve the rotational symmetry of the $E_z(\mathbf{r}_\perp, z_d) = \delta(\mathbf{r}_\perp)$ distribution and we generally observe a donought-shaped profile. We can understand the node at the origin by inspecting Eq. (5) for rotationally symmetric profiles $F_s(\mathbf{k}_\perp)$ and $F_p(\mathbf{k}_\perp)$. In this case we have $A_{x,y}(\mathbf{r}_\perp) = -A_{x,y}(-\mathbf{r}_\perp)$ and $A_{x,y}(0)$ must thus equal 0. AD values for the donought-shaped spots are given in Table 1 and, as previously, the narrower profiles are obtained for opening angles of 60 and 90 degrees.

We observe that the exact shape of the SNOM image of the delta-functions depends on the nature of the TF. The shape is generally not of Gaussian form and this complicates the interpretation of an experimentally obtained SNOM image. In the case where the *x*-component of the field is a delta-function, we have good one-to-one correspondence between delta-function and detected spot for β = 30°, but when β = 90° the two weaker spots neighboring the main spot can be mistaken for near-field point sources of weaker intensity. When the *z*-component of the field is a delta-function the approximation of the images to Gaussian-shaped spots becomes even more ambiguous. We conclude that, even when the concept of resolution is introduced, the identification of the SNOM image as the local electric field intensity is generally incorrect.

If we ignore the non-Gaussian shape of the imaged spots, we can roughly define the resolution of a SNOM tip as twice the AD values given in Table 1. We note that the resolution limit defined here is related to the shape of the tip and is independent of noise.

The definition of resolution described above is meaningful when imaging optical field distributions limited in real space. However, when imaging plane waves that can be considered point sources in $k$-space, the concept of resolution should be treated differently:

In previous works[24,25] the resolution of SNOM setups was determined by imaging counter propagating waves. Two such waves of identical spatial frequency and amplitude propagating along the $x$-axis and decaying along the $z$-axis result in a standing wave pattern of the form $\cos(k_x x)\exp(-|k_z|z)$ with $k_z = \sqrt{k_0^2 - k_x^2}$, where $k_x > k_0$. The intensity pattern features an image contrast $C_i = (I_{max} - I_{min})/(I_{max} + I_{min})$ of 1, which, without background noise, would be preserved in the SNOM image. However, for increasing $k_x$ and thus increasing $|k_z|$ the calculated TFs show us that the standing wave pattern intensity in the image decays near-exponentially, and when it approaches the background noise floor, the image contrast will deteriorate and the limit of the spatial resolution has been reached. It should be stressed that, contrary to the previous case, the resolution limit when imaging a point-source in $k$-space is due to background noise: The coupling coefficient drops exponentially for increasing $k$, but it is never zero. Even though the fiber tip has a finite size, the image contrast would, without noise, be equal to 1 regardless of the value of the spatial frequency.

Keeping the illumination intensity fixed and choosing a noise floor, resolutions of the various tips for the two polarizations can be extracted directly from Fig. 3 as the intersection of the TF curves with the noise floor. Resolutions of the tips described in the figure are given in Table 2 for (arbitrarily chosen) noise floors of 1 pW and 100 fW.

Obviously, the values of the resolving power presented here depend directly on the illumination power and the background noise floor, but relative comparison between the resolutions of various fiber tips can still be made: First, the resolutions for *s*-polarization are slightly better than those for *p*-polarized light. And second, we observe that the resolution is generally improved with ~ 40 % when the noise floor is decreased from 1 pW to 100 fW.

Whether we consider point sources in real space or in *k*-space, the best resolution is not obtained for β = 30° as one might expect but rather for the large opening angle β = 90°. The data show that if one is interested in improving the resolution of a SNOM microscope, one should not aim at producing a fiber tip with smallest opening angle possible.

### *4.2.     Effective plane of detection approximation*

In an effective plane of detection (EPD) approximation[21] the coupling of the near-field to the fiber tip with apex in the detection plane is proportional to the value of the free-space field produced in absence of tip a certain height *h* above the detection plane. From $k_z = \sqrt{k_0^2 - k^2}$ we have that $|k_z| \to k$ for $k >> k_0$, so an evanescent wave with spatial dependence of the form $\exp(ikr - |k_z|z)$ in the reference plane will have decayed by a factor ~ exp(−*kh*) at the height *h*. We recognize this exponential decay as the near-linear regime in Fig. 3 of the TF curves plotted using semi-logarithmic scale for $k/k_0 > 2$. The coupling of the near-field to the fiber tip in this regime can thus be interpreted as coupling at an effective plane of detection positioned a distance *h* above the apex.

However, for $k/k_0 < 2$ the TF curves are not linear. The EPD height *h* can still be defined but it should then be a function of *k*: For a given *k*, we solve the two equations $S(k) = a\exp(2ik_z(k)h)$ and $S(k + \Delta k) = a\exp(2ik_z(k + \Delta k)h)$ for the parameters *a* and *h* when $\Delta k \to 0$ (the factor 2 is included as we fit to power, not field strength). We obtain the function $h(k)$, illustrated in Fig. 5, representing the local EPD height, near the spatial frequency $k$. The function $h(k)$ is highly non-uniform when $k/k_0 < 2$ for both polarizations, and for p-polarized light we observe negative $h$ values for opening angles of 30 and 60 degrees in the intervals where the TFs have a positive slope. When $k/k_0 > 2$, the curves are nearly independent of opening angle and, while still not constant, they vary much less than in the previous regime. The curves are generally slowly decreasing with average value of ~ 125 nm for both polarizations in the interval $2 < k/k_0 < 3.5$. We note that this value agrees well with the EPD height determined when comparing the experimental and modeled data of Ref. 17.

The EPD approximation which assumes a $k$-independent effective plane of detection is intuitively appealing, but Fig. 5 reveals that the approximation is not ideal as it only holds for spatial frequency components of $k/k_0 > 2$. This means that it can be used only when prior knowledge of the field distribution is available, knowledge that allows us to rule out the existence of components with $k/k_0 < 2$. Furthermore, even when restricting the approximation to $k/k_0 > 2$, the effective detection height $h$ is still not exactly constant but continues to decrease with $k$.

In many experiments however, as when imaging the guided modes in photonic crystal structures[7] or the propagation of surface plasmon polaritons, the spectrum of the optical field distribution is limited to a very narrow interval in $k$-space and light of only one polarization is present. In this case the SNOM image does indeed reflect the intensity distribution in an effective detection plane and, if knowledge of the average spatial frequency is available, the $z$-coordinate of this plane can be determined from Fig. 5.

## *4.3.    Fiber tip sharpness*

To examine the importance of employing a sharp fiber tip, we investigated the TFs of fiber tips of opening angle $\beta = 90°$, having their outer end cut off as illustrated in Fig. 6. The detection plane is then raised to the end of the cut-off tip. The TFs for varying tip cut-off diameters $d_t$ are illustrated in Fig. 7.

For $d_t = 0$ μm, nothing has been cut off and the TF is that of Fig. 3b for $\beta = 90°$. For $d_t > 0$ μm the curves feature a complex oscillatory behavior. However, for the $d_t \leq 4$ μm curves, the average power level is identical to or slightly better than that of the uncut tip. This indicates that a perfectly sharp fiber tip results in a near-linear TF curve, where the EPD approximation may be used. But, surprisingly, a perfectly sharp tip is not necessary to achieve good coupling in the evanescent regime. For example, considering the detection of evanescent waves with relatively low spatial frequencies $1 < k/k_0 < 2.5$, the tip cut with $d_t = 0.25$ μm is notably more efficient than the sharp one and preferential over other cut tips due to a monotonous decay of the transfer function.

However, we observe an overall drop in the power level when $d_t$ is increased beyond 4 μm. Also, for cut tips with $d_t$ above ~ 1 μm we notice an improved coupling of propagating waves with $k/k_0$ below ~ 0.4. The amplification of propagating modes in the SNOM image is inconvenient as they will dominate over the evanescent field of interest. We conclude that even though the tip need not be perfectly sharp, the conical part of the tapered fiber beyond $d_t = 1$ μm improves the near-field image and should not be omitted.

## 5. Discussion

The fact that arbitrarily small field variations are not observed in the SNOM image can be roughly understood by introducing the concept of resolution, where the TF is considered a step-function, constant for $k < 2\pi/r$, where $r$ is the resolution limit, and zero otherwise. The SNOM images of delta-functions, however, demonstrate that the SNOM image does not faithfully depict the optical field intensity and that the concept of resolution cannot alone explain the field-image relation.

From the calculated TFs it is clear that, in the evanescent regime, the curves are better approximated to an exponential decay characteristic for an effective plane of detection above the tip apex than to the constant obtained, if the field was effectively coupled at the very apex. Unfortunately, the EPD approximation is only valid for perfectly sharp fiber tips in certain spatial frequency regimes.

The limitations of the concept of resolution and of the EPD approximation suggest that, to interpret the SNOM image correctly, a more general approach is needed: We can consider the SNOM image detected for a given optical field as the result of operating on the field with a coupling operator $\hat{H}$. If this operator is invertible, we can operate with its inverse $\hat{H}^{-1}$ on the SNOM image to obtain the optical field. From Eq. (5) we observe that, in the general case, the operator $\hat{H}$ is invertible if the complex functions $A_x(\mathbf{r}_\perp)$ and $A_y(\mathbf{r}_\perp)$ are known and we thus require that our SNOM microscope can detect phase and power of the two linearly polarized guided modes $(A_x, A_y)$ individually. If this is possible, the components of the optical field are given by the inverse Fourier transform of Eq. (5):

$$\begin{bmatrix} F_s(k,\theta) \\ F_p(k,\theta) \end{bmatrix} = \frac{1}{4\pi^2} \begin{bmatrix} -\sin\theta/H_s(k) & \cos\theta/H_s(k) \\ \cos\theta/H_p(k) & \sin\theta/H_p(k) \end{bmatrix} \int \begin{bmatrix} A_x(\mathbf{r}_\perp) \\ A_y(\mathbf{r}_\perp) \end{bmatrix} \exp(-i\mathbf{k}_\perp \cdot \mathbf{r}_\perp) d\mathbf{r}_\perp \qquad (7)$$

Equation (7) shows that all information about the vectorial electric field in the reference plane is available if we can measure $(A_x(\mathbf{r}_\perp), A_y(\mathbf{r}_\perp))$ and we have prior knowledge of the functions $H_s(k)$ and $H_p(k)$. If the field is uniform along the *y*-axis, the integrals in Eq. (7) vanish for angles $\theta$ different from 0 (mod $\pi$) and the polarizations are separated. In this particular case we can reconstruct the field without individual measurements of $A_x(\mathbf{r}_\perp)$ and $A_y(\mathbf{r}_\perp)$.

Theoretically, the functions $H_s(k)$ and $H_p(k)$ can be completely arbitrary. In practice there will be noise present and we cannot use the inversion of Eq. (7) for spatial frequencies that couple to power levels below the noise floor. To overcome the limitation to resolution due to noise, a nano-collector with a near-constant amplitude of the TF is preferred: With a near-constant amplitude, a weak propagating background field will not be amplified in the SNOM image compared to field components of high spatial frequency, and the evanescent field of interest can then be studied.

## 6. Conclusion

When working with uncoated fiber tips, the detected SNOM image only represents the optical field intensity distribution in an effective detection plane in a few particular cases. In general, an inversion operation must be performed to reconstruct the optical field. This reconstruction requires exact knowledge of the transfer function of the fiber tip and the capability to measure phase and power of the two orthogonal modes of the single-mode fiber individually.

To extend the achievable resolution, not only good coupling but also a flat amplitude of the transfer function, is preferred. Our simulations indicate that for uncoated fiber tips at the wavelength of 633 nm, a tip having a ~ 90° degree opening angle is the optimal choice. The tip can be sharp, but an equally good coupling is obtained for a cut-off tip with diameter $d_t$ up to 1 µm.

This research is carried out in the framework of Center for Micro-Optical Structures (CEMOST) supported by the Danish Ministry for Science, Technology and Innovation, contract No. 2202-603/40001-97.

## Appendix

Using bra-ket notation, we define $|k,\theta,s\rangle \equiv \exp(i\mathbf{k}_\perp \cdot \mathbf{r}_\perp)\mathbf{e}_\theta$ and $|k,\theta,p\rangle \equiv \exp(i\mathbf{k}_\perp \cdot \mathbf{r}_\perp)\mathbf{e}_k$ and Eq. (4) becomes:

$$\mathbf{E}(\mathbf{r}_\perp, z_d) = \int \left(F_s(k,\theta)|k,\theta,s\rangle + F_p(k,\theta)|k,\theta,p\rangle\right) d\mathbf{k}_\perp . \qquad (8)$$

We now introduce the coupling operator $\hat{H}$ coupling the plane wave $|k,\theta,s\rangle$ to the guided mode $\mathbf{G}^{\pm 1}(r)\exp(\pm i\varphi)$ of angular momentum ±1 of a single-mode fiber. The coupling coefficient is then of the form $\langle \pm 1|\hat{H}|k,\theta,s\rangle$. We then exploit that, due to cylindrical symmetry, the rotation operator $e^{-iJ_z\theta}$ and the coupling operator $\hat{H}$ commute:

$$\langle \pm 1|\hat{H}|k,\theta,s\rangle = \langle \pm 1|\hat{H}e^{-iJ_z\theta}|k,0,s\rangle = e^{\mp i\theta}\langle \pm 1|\hat{H}|k,0,s\rangle . \qquad (9)$$

Similar expressions hold for *p*-polarization. We now define coupling coefficients $H_s^{\pm 1}(k) \equiv i\sqrt{2}\langle \pm 1|\hat{H}|k,0,s\rangle$ and $H_p^{\pm 1}(k) \equiv \sqrt{2}\langle \pm 1|\hat{H}|k,0,p\rangle$. The coupling of the total field to the fiber mode becomes:

$$A^{\pm 1}(\mathbf{r}_\perp) = \int \left( \frac{1}{i\sqrt{2}} H_s^{\pm 1}(k) F_s(k,\theta) + \frac{1}{\sqrt{2}} H_p^{\pm 1}(k) F_p(k,\theta) \right) \exp(\mp i\theta + i\mathbf{k}_\perp \cdot \mathbf{r}_\perp) d\mathbf{k}_\perp . \quad (10)$$

This expression describes the coupling to the guided modes $\mathbf{G}^{\pm 1}(r)\exp(\pm i\varphi)$. However, for a weak index contrast between core and cladding, there exist linear combinations $\mathbf{G}_{x,y}(\mathbf{r}_\perp)$ of the modes $\mathbf{G}^{\pm 1}(r)\exp(\pm i\varphi)$ which to a good approximation are linearly polarized. We are therefore usually more interested in coupling to the modes $\mathbf{G}_{x,y}(\mathbf{r}_\perp)$ and we thus define $|x\rangle = (|1\rangle + |-1\rangle)/\sqrt{2}$ and $|y\rangle = (|1\rangle - |-1\rangle)/(i\sqrt{2})$. Now, as incoming s- and p-polarized light of angle θ = 0 is directed along the y- and x-axis respectively, we have $\langle x|\hat{H}|k,0,s\rangle = 0$ and $\langle y|\hat{H}|k,0,p\rangle = 0$ and obtain the properties $H_s^1 = -H_s^{-1} \equiv H_s$ and $H_p^1 = H_p^{-1} \equiv H_p$ for the coupling coefficients. The coupling of the field to the mode profiles $A_{x,y}\mathbf{G}_{x,y}(\mathbf{r}_\perp)$ is then given by the expression:

$$\begin{bmatrix} A_x(\mathbf{r}_\perp) \\ A_y(\mathbf{r}_\perp) \end{bmatrix} = \int \begin{bmatrix} -\sin\theta & \cos\theta \\ \cos\theta & \sin\theta \end{bmatrix} \begin{bmatrix} H_s(k) F_s(k,\theta) \\ H_p(k) F_p(k,\theta) \end{bmatrix} \exp(i\mathbf{k}_\perp \cdot \mathbf{r}_\perp) d\mathbf{k}_\perp . \quad (11)$$

Fig. 1: (Color online) Illustration of the SNOM measurement setup.

Fig. 2: Unit vectors in cylindrical coordinates.

Fig. 3: (Color online) Transfer functions for opening angles β from 30° to 120° for *s*-polarization (a) and *p*-polarization (b).

Fig. 4: (Color online) SNOM images of near-field delta functions as function of opening angle β. In (a) the *x*-component of the field is a delta-function. In (b) the *z*-component is a delta-function.

Fig. 5: (Color online) Local effective plane of detection height for *s*-polarization (a) and *p*-polarization (b).

Fig. 6: (Color online) Illustration of cut-off tip geometry. The end of the cut-off tip has a diameter $d_t$.

Fig. 7: (Color online) Transfer functions for *p*-polarization for tips of fixed opening angle β of 90° with varying cut-off diameter $d_t$.

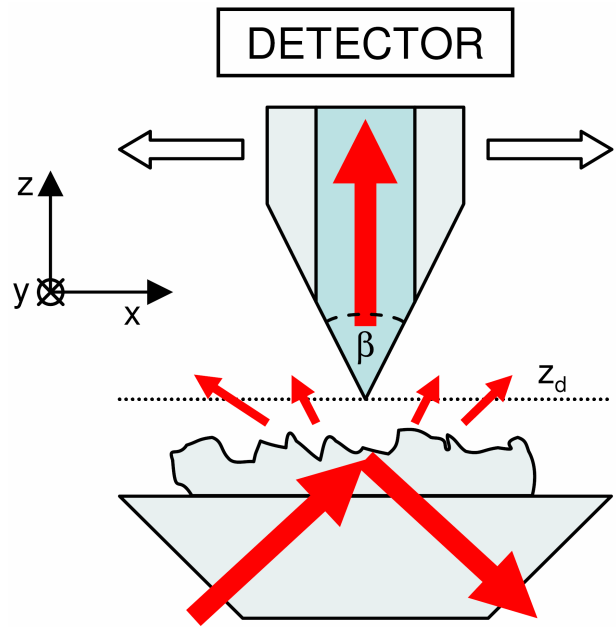

Fig. 1: (Color online) Illustration of the SNOM measurement setup.

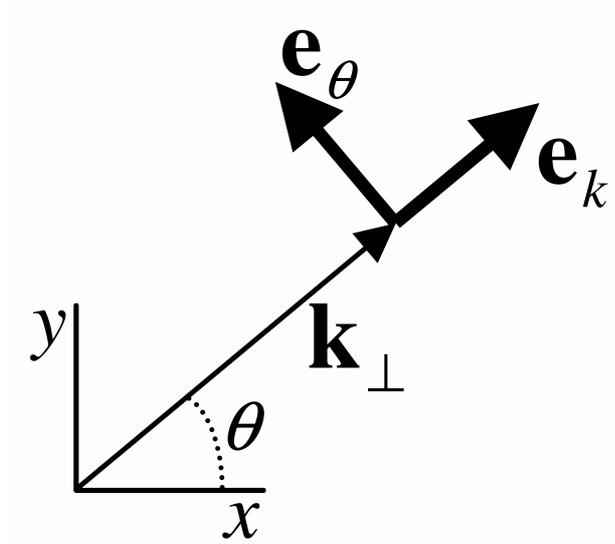

Fig. 2: Unit vectors in cylindrical coordinates.

Note to editor: The figure below should be of two columns width.

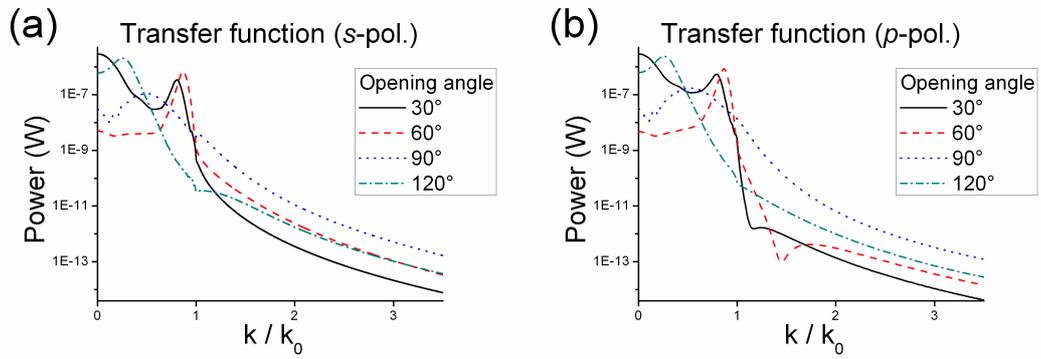

Fig. 3: (Color online) Transfer functions for opening angles β from 30° to 120° for *s*-polarization (a) and *p*-polarization (b).

Note to editor: The figure below should be of two columns width.

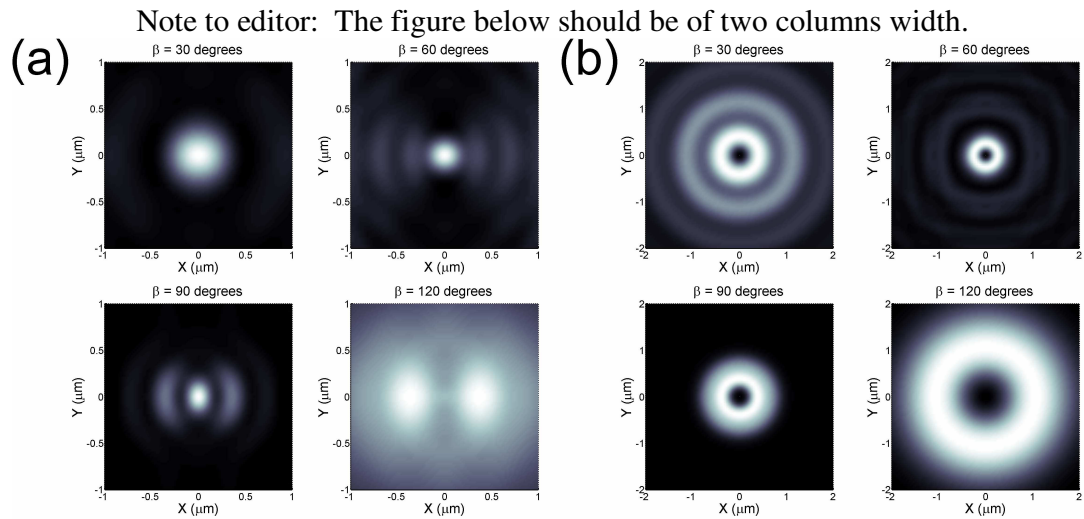

Fig. 4: (Color online) SNOM images of near-field delta functions as function of opening angle β. In (a) the *x*-component of the field is a delta-function. In (b) the *z*-component is a delta-function.

Note to editor: The figure below should be of two columns width.

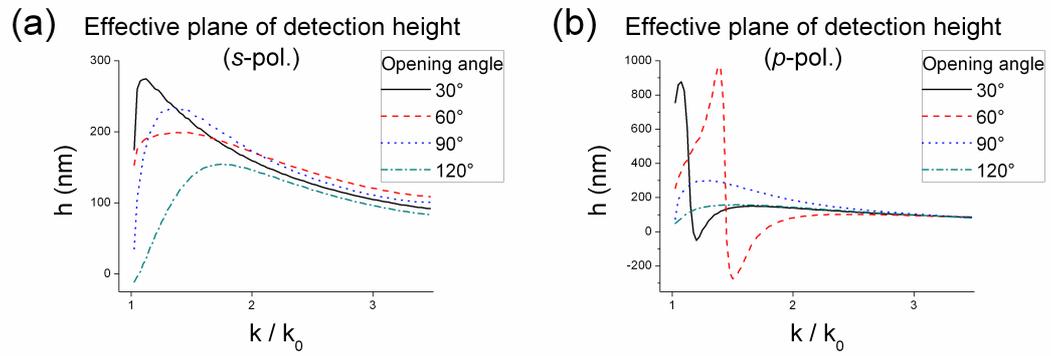

Fig. 5: (Color online) Local effective plane of detection height for *s*-polarization (a) and *p*-polarization (b).

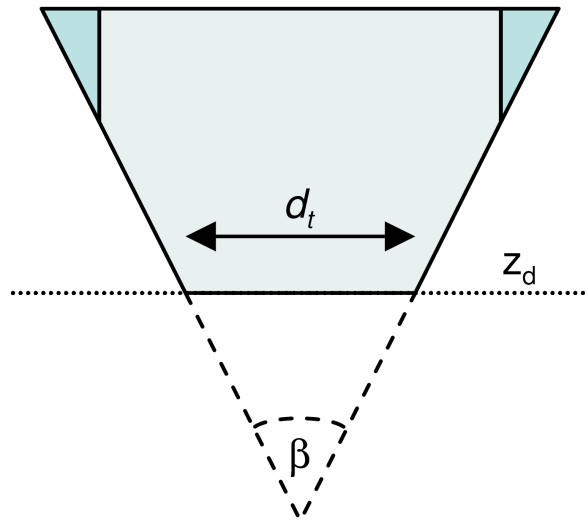

Fig. 6: (Color online) Illustration of cut-off tip geometry. The end of the cut-off tip has a diameter $d_t$.

Note to editor: The figure below should be of two columns width.

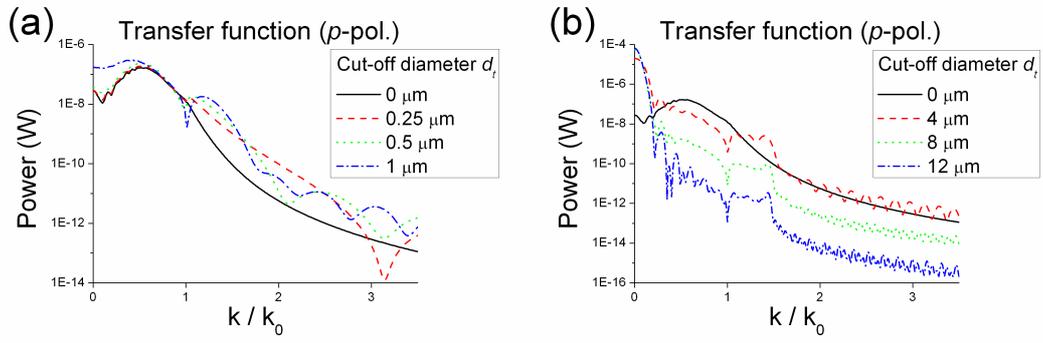

Fig. 7: (Color online) Transfer functions for *p*-polarization for tips of fixed opening angle β of 90° with varying cut-off diameter $d_t$.

**Table 1: Average deviation of images of delta-functions.**

| | $\mathbf{E}_\perp = \delta(\mathbf{r}_\perp)\mathbf{e}_x$ | | $E_z = \delta(\mathbf{r}_\perp)$ |
|---|---|---|---|
| Opening angle β | Average deviation along *x*-axis | Average deviation along *y*-axis | Average deviation |
| 30° | 262 nm | 160 nm | 1175 nm |
| 60° | 1029 nm | 88 nm | 845 nm |
| 90° | 255 nm | 104 nm | 506 nm |
| 120° | 698 nm | 683 nm | 1199 nm |

**Table 2: SNOM tip resolving power.**

| Opening angle β | Noise floor | | | |
|---|---|---|---|---|
| | 1 pW | 100 fW | 1 pW | 100 fW |
| | Resolution (*s*-pol.) | | Resolution (*p*-pol.) | |
| 30° | 365 nm | 264 nm | 437 nm | 300 nm |
| 60° | 286 nm | 209 nm | 471 nm | 251 nm |
| 90° | 219 nm | 156 nm | 248 nm | 180 nm |
| 120° | 289 nm | 205 nm | 317 nm | 222 nm |